%% file: paper.tex
\documentclass[a4paper,USenglish,cleveref]{lipics-v2021}

\hideLIPIcs
\nolinenumbers

\newcommand{\myparagraph}[1]{\subparagraph*{#1}}
\input{pgf_header}

\usepackage{dsfont}
\usepackage{mathbbol}
\usepackage{pgfplots}
\usepackage{caption}
\usepackage{subcaption}
\usepackage{hyperref}
\usepackage{booktabs}
\usepackage[nocompress]{cite}
\crefname{listing}{Algorithm}{Algorithms}

\newcommand{\mytitle}{\texorpdfstring{Brief Announcement:\\Parallel Construction of Bumped Ribbon Retrieval}{Brief Announcement: Parallel Construction of Bumped Ribbon Retrieval}}
\title{\mytitle}
\titlerunning{Parallel Construction of Bumped Ribbon Retrieval}

\author{Matthias Becht}{Karlsruhe Institute of Technology, Germany}{}{}{}
\author{Hans-Peter Lehmann}{Karlsruhe Institute of Technology, Germany}{hans-peter.lehmann@kit.edu}{https://orcid.org/0000-0002-0474-1805}{}
\author{Peter Sanders}{Karlsruhe Institute of Technology, Germany}{sanders@kit.edu}{https://orcid.org/0000-0003-3330-9349}{}

\newcommand{\myauthorrunning}{M. Becht, H.-P. Lehmann, P. Sanders}
\authorrunning{\myauthorrunning}
\Copyright{Matthias Becht, Hans-Peter Lehmann, and Peter Sanders}
\hypersetup{
  colorlinks=true,
  pdftitle={\mytitle},
  pdfauthor={\myauthorrunning},
  pdfsubject={}
}


\supplementdetails[subcategory={Source code}]{Software}{https://github.com/lorenzhs/BuRR/tree/parallel}

\funding{
  \flag[2cm]{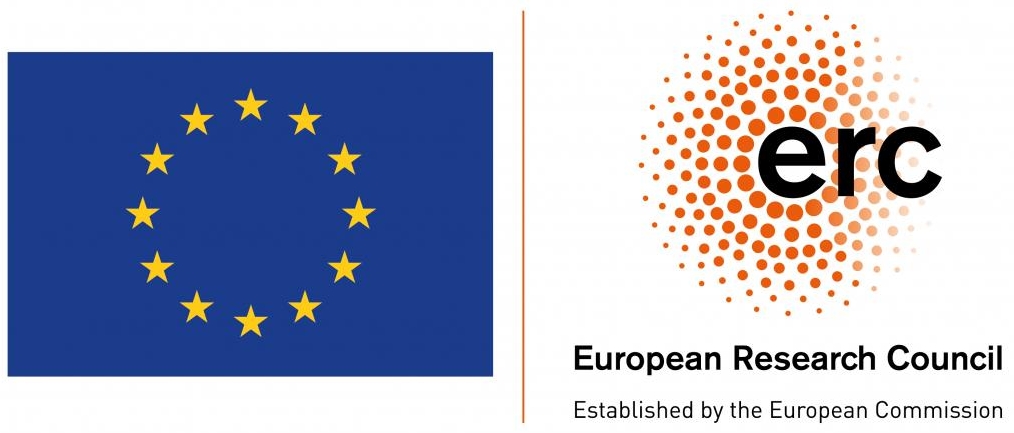}
  This project has received funding from the European Research Council (ERC) under the European Union’s Horizon 2020 research and innovation programme (grant agreement No. 882500).
}

\begin{document}
\maketitle
\input{fig/macros.tex}

\begin{abstract}
  A retrieval data structure stores a static function $f: S \rightarrow \{0, 1\}^r$.
  For all $x \in S$, it returns the $r$-bit value $f(x)$, while for other inputs it may return an arbitrary result.
  The structure cannot answer membership queries, so it does not have to encode $S$.
  The information theoretic space lower bound for arbitrary inputs is $r|S|$ bits.
  Retrieval data structures have widespread applications.
  They can be used as an approximate membership filter for $S$ by storing fingerprints of the keys in $S$, where they are faster and more space efficient than Bloom filters.
  They can also be used as a basic building block of succinct data structures like perfect hash functions.

  Bumped Ribbon Retrieval (BuRR) [Dillinger et al., SEA'22] is a recently developed retrieval data structure that is fast to construct with a space overhead of less than 1\%.
  The idea is to solve a nearly diagonal system of linear equations to determine a matrix that, multiplied with the hash of each key, gives the desired output values.
  During solving, BuRR might \emph{bump} lines of the equation system to another layer of the same data structure.
  While the paper describes a simple parallel construction based on bumping the keys on thread boundaries, it does not give an implementation.
  In this brief announcement, we now fill this gap.

  Our parallel implementation is transparent to the queries.
  It achieves a speedup of \speedupMax{} on 32 cores for 8-bit filters.
  The additional space overhead is \bytesOverheadPerThread{} bytes per thread, or \bytesOverheadPerThread{} slots.
  This matches \bytesOverheadPercentage{}\% of the total space consumption when constructing with 1 billion input keys.
  A large portion of the construction time is spent on parallel sorting.
\end{abstract}
\newpage

\section{Introduction}
A retrieval data structure \cite{majewski1996family} can store a static function $f: S \rightarrow \{0, 1\}^r$.
For all $x \in S$, it returns the $r$-bit value $f(x)$, while for all other inputs it may return an arbitrary result.
In contrast to a hash table, a retrieval data structure cannot answer membership queries (``is $x \in S$?'').
This makes it possible to store $f$ without encoding $S$.
The information theoretic space lower bound is $r|S|$ bits.

The recently developed Bumped Ribbon Retrieval (BuRR) \cite{dillinger2022fast} is a very fast and space-efficient approach to the retrieval problem.
In its most compact configuration, it has a space overhead of below 1\% to the information theoretic space lower bound.
The idea is to solve a nearly diagonal system of linear equations to determine a matrix that the hash of queried keys can be multiplied with to get the stored value.
Each equation corresponds to one input key.
During solving, BuRR might \emph{bump} lines (or their corresponding keys) to another layer of the same data structure.
We describe BuRR in more detail in \cref{s:prelims}.

\myparagraph{Applications.}
Retrieval data structures have widespread applications.
They can be used as an approximate membership filter by storing fingerprints of the keys in $S$, where they are faster and more space efficient than classical Bloom filters \cite{bloom1970space}.
The key-value database ``rocksdb'' \cite{facebook2021rocksdb} developed by Facebook uses BuRR in its log-structured merge tree to filter out external memory access operations.

Retrieval data structures can also be used as a basic building block of succinct data structures like perfect hash functions.
A minimal perfect hash function maps a set $S$ of keys to the first $|S|$ integers without collisions.
SicHash \cite{lehmann2023sichash} hashes each key to a small number of candidate positions.
It uses BuRR to store a candidate to use for each key such that no keys collide.
A large share of the SicHash construction time is spent constructing these retrieval data structures.
ShockHash \cite{lehmann2024shockhash,lehmann2023bipartite} uses a similar idea.
LeMonHash \cite{ferragina2023learned} is a \emph{monotone} minimal perfect hash function that maps each input key to its rank.
It determines an estimate for the ranks using a learned index data structure and solves collisions in that rank estimate by storing local ranks in a collection of BuRR retrieval data structures.
Refs. \cite{lehmann2023sichash,ferragina2023learned,lehmann2024shockhash} specifically state that they need a parallel BuRR implementation to parallelize their approach.

\myparagraph{Parallelization.}
BuRR can be constructed in parallel, as described in the original paper \cite{dillinger2022fast}.
However, their implementation is only sequential, which limits the construction throughput of applications using BuRR.
In this brief announcement, we implement the parallel construction, which we describe in \cref{s:parallel}.
We then evaluate the performance in \cref{s:experiments}.

\section{Preliminaries}\label{s:prelims}
\begin{figure}[t]
  \begin{subfigure}[t]{0.585\textwidth}
    \centering
    \includegraphics[scale=0.9,page=1]{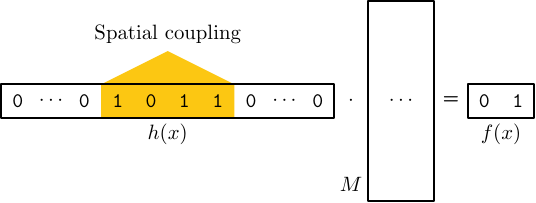}
    \caption{
      The query operation multiplies a hash of the input key with the matrix $M$, which is the main data structure.
    }
    \label{prelim:fig:ribbonQuery}
  \end{subfigure}
  \hfill
  \begin{subfigure}[t]{0.375\textwidth}
    \centering
    \includegraphics[scale=0.9,page=2]{fig/ribbon}
      \caption{The equation system to solve during construction has a small band (\emph{ribbon}) in which there can be 1-bits.}
    \label{prelim:fig:ribbonConstruction}
  \end{subfigure}
  \caption{Illustrations of Bumped Ribbon Retrieval (BuRR) \cite{dillinger2022fast} with $r=2$ bits. Simplified here to ignore bumping.}
  \label{prelim:fig:ribbon}
\end{figure}

Ribbon Retrieval \cite{dietzfelbinger2019efficient} stores a table of $r$-bit values.
In the following, we call each table row a \emph{slot}.
The 1-bits in the hash of a key determine which slots should be \texttt{XOR}ed together to give the $r$-bit output value for that key.
Construction then boils down to solving a system of linear equations over $\{0, 1\}^r$ to determine the table.
Ribbon Retrieval uses hash functions with \emph{spacial coupling}:
Only values within a certain \emph{ribbon width} can contain 1-bits.
This leads to the significant advantage that the equation system is almost diagonal, which makes it very efficient to solve using Gaussian elimination.
We illustrate this in \cref{prelim:fig:ribbon}.
Additionally, the use of spatial coupling makes the queries very simple and cache local.
In particular, for $r=1$, the queries boil down to \texttt{AND}ing a segment of the precomputed table with the hash of the input key and reporting the parity of the result.

\emph{Bumped} Ribbon Retrieval (BuRR) \cite{dillinger2022fast} enhances the idea with a concept called \emph{bumping}.
If BuRR encounters an equation that would prevent successful solving, it removes some of the equations from the equation system.
The corresponding keys are then handled in another layer of the same data structure.
To bump keys, BuRR divides the keys into small buckets.
For each bucket, it stores a \emph{threshold} value that indicates which keys should be bumped.
There are several ways of storing the thresholds that provide a trade-off between performance and space overhead.
The most important ones are \emph{2-bit} and \emph{$1^+$-bit}.
For 2-bit thresholds, there are four pre-determined values, one of them bumping the entire bucket and one bumping no keys.
$1^+$-bit thresholds build on the observation that most buckets do not have to bump keys.
The main threshold is just a single bit indicating whether something was bumped.
Only if something was bumped, we store the exact threshold in a small hash table.

For additional space savings, BuRR \emph{overloads} the equation system with more equations than can fit.
While this causes more bumping, it increases the utilization of the table.
Therefore, bumping is an integral part of the design.
For a more detailed explanation, refer to the original paper \cite{dillinger2022fast}.

\section{Parallelization}\label{s:parallel}
For constructing any retrieval data structure in parallel, a naive approach is to use partitioning.
The input set can be hashed to independent partitions that can then be constructed in parallel.
However, this introduces query overhead because every query then needs to check which partition to access.
Additionally, it duplicates constant space overhead for every additional partition, or, respectively, for every thread that is used for construction.

BuRR can be constructed in parallel using a much more elegant approach, as mentioned in the original paper \cite{dillinger2022fast}.
To parallelize BuRR, we can simply divide the equation system into one partition per thread.
To make the partitions independent, we bump keys in the buckets that are located on a thread boundary.
Because BuRR uses the bumping mechanism anyway, the parallelization is transparent to the query algorithm.
While the BuRR paper mentions how to parallelize the construction, it does not give an implementation.
In this brief announcement, we close this gap.

\begin{figure}[t]
  \centering
  \includegraphics[scale=0.9,page=1]{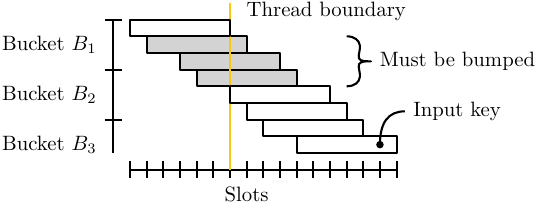}
  \caption{Illustrations of the keys that have to be bumped around the thread boundaries. Note that keys may use the slots in the gap between two threads, they just cannot overlap the thread boundary.}
  \label{prelim:fig:bumping}
\end{figure}

\myparagraph{Bumping.}
In order to divide the equation system into partitions, we must bump at least $w - 1$ slots at the thread boundaries, where $w$ is the ribbon width.
This ensures that the equations of neighboring threads are completely independent.
Note that bumping does not mean that there are unused slots at the thread boundaries.
The slots can still all store payload data, there are just no keys \emph{overlapping} the thread boundaries.
We illustrate this in \cref{prelim:fig:bumping}.
In the illustration, bucket $B_1$ needs to be bumped completely because its last key (with the smallest threshold) needs to be bumped.
Bucket $B_2$ can bump its keys partially.

The last layer of the data structure cannot be processed in parallel since we cannot bump any keys there.
However, note that this base case takes only a negligible part of the overall construction time.

\myparagraph{Minimum Buckets per Thread.}
The additional bumped keys at the thread boundaries introduce some space overhead at the lower layers of the data structure.
To reduce the number of bumped keys, it is useful to set a minimum number of buckets to be processed by each thread, so that fewer threads are used for small input sizes (or in deeper layers of the data structure).
This reduces the space overhead and does not incur a significant performance penalty if chosen correctly since there is also a certain overhead from starting each thread.
In the following, we use the \emph{minbpt} variable to indicate the minimum number of buckets that each thread needs to have.

\myparagraph{Selecting Cut Points.}
We also implement several strategies for small improvements in the space overhead by searching for more appropriate cuts that do not cause as much bumping.
One possible strategy (\emph{minbump}) is to search for the bucket in which the fewest keys have to be directly bumped in order to create a gap in the system of equations.
Another option (\emph{maxprev}) is to search for a bucket which has the most keys in the $w - 1$ slots \emph{before} this bucket, where $w$ is the ribbon width.
The intuition behind this is that we want keys to fill the gap that is created.
The last variant we tried (\emph{diff}) is a combination of both which tries to minimize the difference between the number of directly bumped keys and the number of keys in the $w - 1$ slots before the bucket.
For each of these variants, the number of buckets to be searched can be configured.

\section{Experiments}\label{s:experiments}
In this section, we give our detailed experimental evaluation.
We explain the space overhead in \cref{ss:spaceOverhead} and the scaling in the number of threads in \cref{ss:scaling}.

We run our experiments on an AMD EPYC 7551P with 32 cores (64 hardware threads (HT)) and a base clock speed of 2.0 GHz.
The machine runs Ubuntu 22.04 with Linux 5.15.0.
We use the GNU C++ compiler version 11.4.0 with optimization flags \texttt{-O3 -march=native}.
Our experiments use BuRR as an 8-bit filter, so each key needs about one byte of space.
Our bucket size is 128 keys.
Furthermore, we use BuRR with a depth of 4 layers with parallel insertion before falling back to the single-threaded base-case.
Because the last layer cannot use bumping and needs to be scaled up if its equations cannot be solved, a small depth can lead to significant noise in the achieved space consumption.
These jumps are not a problem in practice because the last layer is small.
However, in our experiments measuring the space overhead in absolute numbers per thread, this is relevant.
For sorting the hashes by their insertion position, we use parallel IPS$^2$Ra \cite{axtmann2020engineering}.
Our implementation is available on GitHub: \url{https://github.com/lorenzhs/BuRR/tree/parallel}.

\subsection{Space Overhead} \label{ss:spaceOverhead}
We now give the space overhead for different threshold modes, different values of \emph{minbpt}, and different search strategies.
In the plots in this section, we give the space overhead per \emph{additional} thread.
We do this because for $t$ threads, there are only $t-1$ cut points.

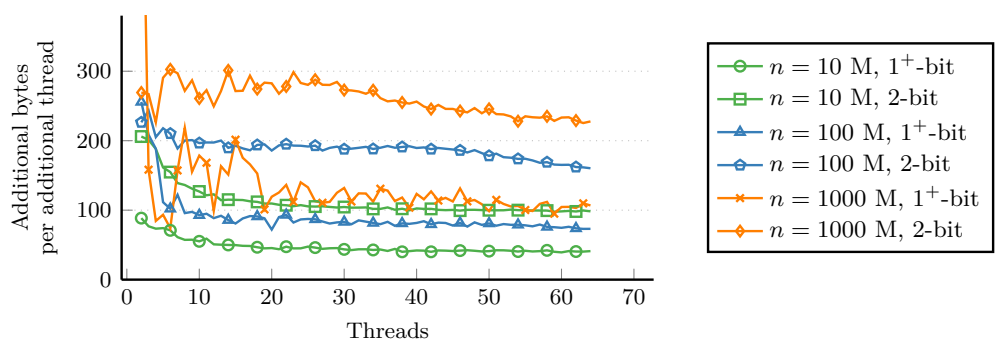
\begin{figure}[t]
    \centering
    \input{fig/spaceoverhead}
    \caption{
      Space overhead in bytes with \emph{minbpt}=1000 with $1^+$-bit and 2-bit thresholds.
    }
    \label{fig:spaceoverhead}
\end{figure}

\myparagraph{Threshold Storage.}
\Cref{fig:spaceoverhead} shows the space overhead in bytes for different input sizes and thresholds, with the minimum buckets per thread (\emph{minbpt}) set to 1000.
Overall, we get a space overhead of about \bytesOverheadPerThread{} bytes for $1^+$-bit thresholds and 100 million keys.
Because we use BuRR as an 8-bit filter, this corresponds to \bytesOverheadPerThread{} slots lost for each additional thread.
Given that the ribbon width is 64 and that we have cuts in all 4 layers, this space overhead is expected.
It indicates that our parallel implementation indeed uses the space left by the bumped keys instead of just creating a section of unused slots (see also \cref{prelim:fig:bumping}).
Note also that bumping complete buckets would cause an overhead of 128 bytes for each layer, so we usually do not bump complete buckets.
The space overhead when using 2-bit thresholds is higher because they are less granular, so they need to bump more keys.
For $1^+$-bit thresholds, almost every thread boundary leads to a hash table entry.
However, this additional space consumption is quickly compensated for by the more granular thresholds, making better use of the slots.
The fact that more input keys have more space overhead is due to the \emph{minbpt} parameter.
With a smaller input set, the lower layers use fewer threads and therefore fewer cut points on which to bump keys.
The noise in the measurements with 1 billion keys can be explained by the fact that we only use a depth of 4 layers.
As mentioned before, having to scale the last layer introduces a tiny \emph{relative} overhead, but can cause rather high \emph{absolute} overhead, which is what we measure here.

\myparagraph{Values of \emph{minbpt}.}

\begin{figure}[t]
    \centering
    \begin{tikzpicture}
        \ref*{legendCombined1}
    \end{tikzpicture}
    \vspace{1mm}

    \begin{subfigure}[t]{0.48\textwidth}
        \input{fig/combined1bit}
        \caption{$1^+$-bit thresholds.}
        \label{fig:combined1bit}
    \end{subfigure}%
    \hfill
    \begin{subfigure}[t]{0.48\textwidth}
        \input{fig/combined2bit}
        \caption{2-bit thresholds.}
        \label{fig:combined2bit}
    \end{subfigure}

    \caption{Space overhead in bytes versus 32-thread speedup for different values of \emph{minbpt}.}
    \label{fig:combined}
\end{figure}
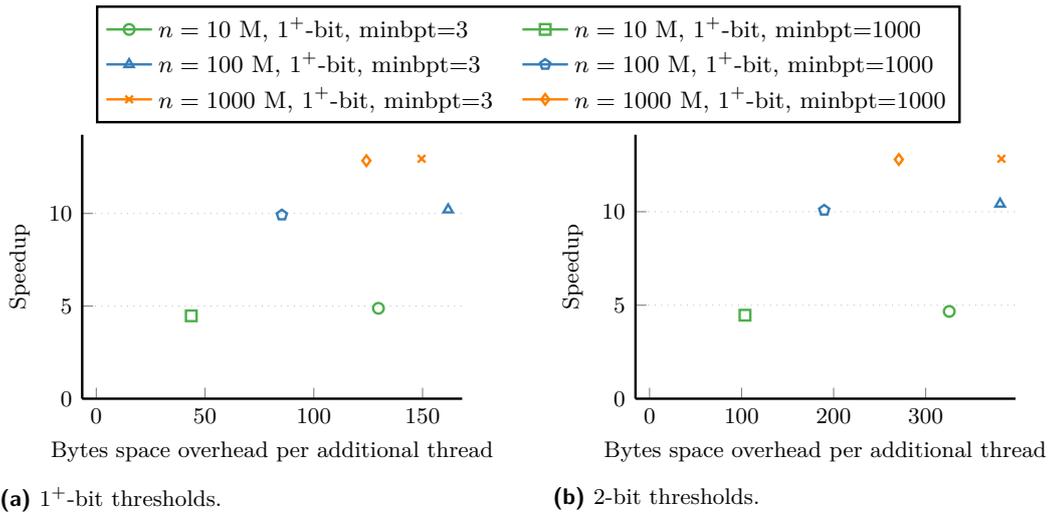

\Cref{fig:combined} gives a comparison of the space overhead and 32-thread speedup with different values for the minimum number of buckets per thread for $1^+$-bit and 2-bit thresholds, respectively.
As would be expected, the space overhead decreases with a higher value of \emph{minbpt} since we get fewer cuts.
Our measurements indicate that \emph{minbpt} can be rather large before we get any performance penalty.
In fact, we only get a small penalty for 2-bit thresholds when selecting \emph{minbpt}=1000.
It is therefore useful to make the value large to reduce the space overhead.
Note that the ideal value for \emph{minbpt} depends on the specific machine being used since a significant factor is the overhead from starting threads.

\subparagraph*{Selecting Cut Points.}

\begin{figure}[t]
    \centering
    \input{fig/search}
    \caption{Space overhead in bytes with different search strategies for the cut points. Uses 100 million keys, 32 threads, \emph{minbpt}=100, and a search range of 50.}
    \label{fig:search}
\end{figure}
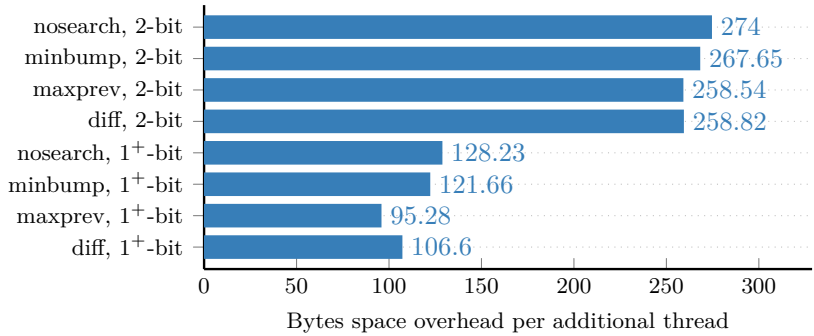

\Cref{fig:search} shows a comparison of the space overhead for different search strategies when using 32 threads and a search range of 50.
The \emph{maxprev} and \emph{diff} strategies slightly improve the space overhead compared to simply cutting at predetermined positions.
The maximum amount of space saved is about \spaceSavingsMaxprev{} bytes per additional thread for $1^+$-bit thresholds and \emph{maxprev} search.
While it reduces the absolute values of the overhead noticeably, this is not significant compared to the millions of bytes of total storage space.
We conclude that smarter search strategies are not worth the additional complexity during construction.
A more reliable way to reduce the space usage at the cost of a slight increase in construction time, is to increase the value of \emph{minbpt}.

\subsection{Scaling} \label{ss:scaling}

\begin{figure}[t]
    \input{fig/speedup}
    \caption{Speedup with \emph{minbpt}=1000 with $1^+$-bit and 2-bit thresholds.}
    \label{fig:speedup}
\end{figure}
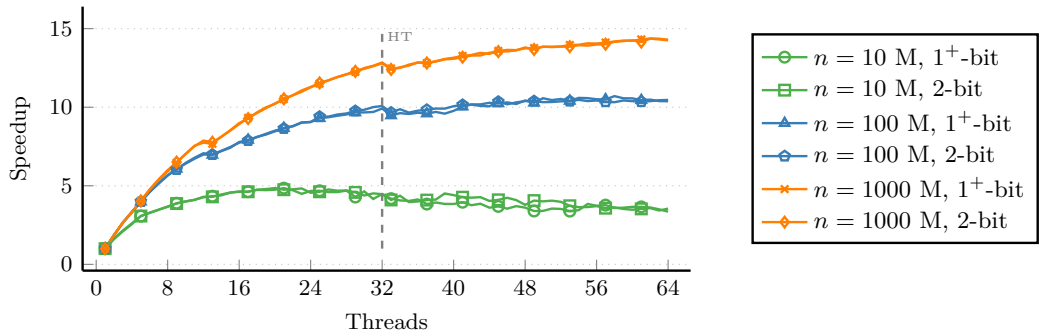

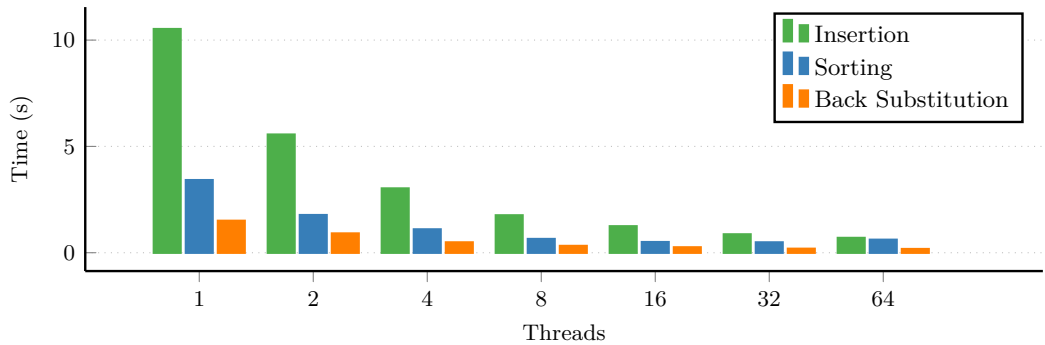
\begin{figure}[t]
    \input{fig/breakdown}
    \caption{Breakdown of runtime for 100 million keys with \emph{minbpt}=1000 and 2-bit thresholds.}
    \label{fig:breakdown}
\end{figure}

\Cref{fig:speedup} gives the speedup for 2-bit and $1^+$-bit thresholds when constructing with 10 million, 100 million, and 1 billion keys.
A larger number of input keys allows for a significantly higher speedup.
This is because, for 10 million keys, the choice of \emph{minbpt} means that we already construct the second layer with a reduced number of threads.
For 1 billion keys, we get a speedup of about \speedupMax{} for 32 cores (64 threads).
\Cref{fig:breakdown} shows the bottleneck of the parallelization.
While the insertion profits from more threads, the sorting step does not scale well for more than 8 threads.
With 64 threads, the sorting step then takes close to half of the construction time.
Given that we already use the highly optimized parallel sorter IPS$^2$Ra \cite{axtmann2020engineering}, we assume that it is hard to achieve even better speedups.
However, our parallel implementation gives significant speedups that accelerate applications using BuRR while being transparent to the queries.

\section{Conclusion and Future Work}\label{s:conclusion}
In this brief announcement, we have parallelized the BuRR retrieval data structure \cite{dillinger2022fast} using the approach outlined in the original paper.
The idea is to use the existing bumping mechanism to ensure that the partition handled by each thread is independent.
In contrast to a naive partitioning layer on top of the data structure, this approach is transparent to the queries.
Our implementation shows a speedup of \speedupMax{} when constructing with 32 cores.
Because the per-thread space overheads are quite small, this approach seems feasible for a GPU construction in future work.

\bibliography{paper}

\end{document}

%% file: pgf_header.tex
\usepackage{pgfplots}
\pgfplotsset{compat=newest}

\usepgfplotslibrary{groupplots}
\pgfplotsset{every axis/.style={scale only axis}}

\definecolor{colorPlotGreen}{HTML}{4DAF4A}
\definecolor{colorPlotOrange}{HTML}{FF7F00}
\definecolor{colorPlotBlue}{HTML}{377EB8}
\definecolor{colorPlotGray}{HTML}{444444}
\definecolor{colorPlotBrown}{HTML}{A65628}
\definecolor{colorPlotPurple}{HTML}{984EA3}
\definecolor{colorPlotPink}{HTML}{F781BF}

\pgfplotscreateplotcyclelist{myColorList}{%
  colorPlotGreen,mark=o\\%
  colorPlotBlue,mark=square\\%
  colorPlotOrange,mark=triangle\\%
  colorPlotGray,mark=pentagon\\%
  colorPlotPurple,mark=x\\%
  colorPlotBrown,mark=diamond\\%
  colorPlotPink,mark=flippedTriangle\\%
}
\pgfdeclareplotmark{flippedTriangle}{%
  \pgfpathmoveto{\pgfqpointpolar{-90}{1.2\pgfplotmarksize}}%
  \pgfpathlineto{\pgfqpointpolar{30}{1.2\pgfplotmarksize}}%
  \pgfpathlineto{\pgfqpointpolar{150}{1.2\pgfplotmarksize}}%
  \pgfpathclose%
  \pgfusepath{stroke}
}
\pgfplotsset{
  mark repeat*/.style={
    scatter,
    scatter src=x,
    scatter/@pre marker code/.code={
      \pgfmathtruncatemacro\usemark{
        or(mod(\coordindex,#1)==0, (\coordindex==(\numcoords-1))
      }
      \ifnum\usemark=0
        \pgfplotsset{mark=none}
      \fi
    },
    scatter/@post marker code/.code={}
  },
  major grid style={thin,dotted},
  minor grid style={thin,dotted},
  ymajorgrids,
  yminorgrids,
  every axis/.append style={
    line width=0.9pt,
    tick style={
      line cap=round,
      thin,
      major tick length=4pt,
      minor tick length=2pt,
    },
    mark options={solid},
  },
  legend cell align=left,
  legend style={
    line width=0.9pt,
    /tikz/every even column/.append style={column sep=3mm,black},
    /tikz/every odd column/.append style={black},
    mark options={solid},
  },
  legend style={font=\small},
  title style={yshift=-2pt},
  enlarge x limits=0.04,
  every tick label/.append style={font=\footnotesize},
  every axis label/.append style={font=\small},
  every axis y label/.append style={yshift=-1ex},
  /pgf/number format/1000 sep={},
  axis lines*=left,
  xlabel near ticks,
  ylabel near ticks,
  axis lines*=left,
  label style={font=\footnotesize},
  tick label style={font=\footnotesize},
  cycle list name=myColorList,
}

%% file: fig/macros.tex
\def\speedupMax{14}

\def\bytesOverheadPercentage{0.0007}

\def\bytesOverheadPerThread{105}

\def\spaceOverheadNosearch{128.227956989119}

\def\spaceOverheadMaxprev{95.2838709676458}

\def\spaceSavingsMaxprev{32}

%% file: fig/spaceoverhead.tex
\begin{tikzpicture}
    \begin{axis}[
      xlabel={Threads},
      ylabel={Additional bytes\\per additional thread},
      ylabel style={align=center},
      legend style={at={(1.1, 0.5)},anchor=west},
      xmax=70,
      ymax=380,
      ymin=0,
      width=7cm,
      height=3.5cm,
      mark repeat=4,
    ]
    \addplot+[colorPlotGreen] coordinates { (2,88.5333) (3,76.8) (4,73.3556) (5,74.1333) (6,70.9733) (7,61.1556) (8,57.2571) (9,57.4667) (10,55.1481) (11,58.98) (12,50.7939) (13,50.2778) (14,49.9333) (15,49.7) (16,48.6667) (17,48.4625) (18,46.6353) (19,45.0407) (20,45.4491) (21,43.7767) (22,47.4032) (23,44.997) (24,45.0493) (25,47.525) (26,46.1173) (27,44.3282) (28,45.1654) (29,45.4048) (30,43.5701) (31,42.5333) (32,43.6731) (33,43.7021) (34,42.7131) (35,43.3431) (36,41.2895) (37,44.2148) (38,40.0541) (39,41.0088) (40,41.988) (41,41.735) (42,40.0634) (43,41.9952) (44,41.7302) (45,41.1788) (46,41.7941) (47,42.3188) (48,41.5504) (49,40.5903) (50,41.0422) (51,42.272) (52,42.2039) (53,41.5654) (54,40.4415) (55,41.5395) (56,40.543) (57,41.0333) (58,42.152) (59,40.108) (60,39.2203) (61,41.0978) (62,40.4372) (63,40.3613) (64,41.055) };
    \addlegendentry{$n=10$ M, $1^+$-bit}
    \addplot+[colorPlotGreen] coordinates { (2,206.0) (3,201.167) (4,188.2) (5,161.45) (6,154.8) (7,138.956) (8,136.238) (9,130.958) (10,126.844) (11,121.433) (12,122.818) (13,114.378) (14,115.118) (15,114.829) (16,114.578) (17,112.762) (18,111.91) (19,109.496) (20,109.582) (21,107.743) (22,107.086) (23,105.712) (24,107.243) (25,105.458) (26,105.168) (27,105.562) (28,104.662) (29,103.826) (30,104.517) (31,103.6) (32,103.701) (33,103.9375) (34,101.826) (35,101.849) (36,103.945) (37,100.581) (38,102.023) (39,99.7895) (40,102.492) (41,101.852) (42,101.868) (43,100.46) (44,100.019) (45,100.365) (46,99.3704) (47,100.467) (48,100.515) (49,100.107) (50,99.9034) (51,99.7907) (52,99.8444) (53,100.795) (54,100.025) (55,99.758) (56,99.577) (57,99.4821) (58,99.3836) (59,97.8172) (60,98.2644) (61,98.9044) (62,98.1202) (63,99.5075) (64,98.3344) };
    \addlegendentry{$n=10$ M, 2-bit}
    \addplot+[colorPlotBlue] coordinates { (2,255.867) (3,208.5) (4,188.444) (5,111.45) (6,102.053) (7,122.2) (8,95.7714) (9,97.55) (10,92.837) (11,94.84) (12,88.5333) (13,91.9222) (14,85.7538) (15,80.8095) (16,88.0667) (17,91.475) (18,90.9608) (19,88.2556) (20,72.0386) (21,90.03) (22,93.1524) (23,82.2) (24,86.658) (25,87.3583) (26,86.84) (27,83.6897) (28,80.7605) (29,82.3952) (30,83.0874) (31,79.7133) (32,85.3161) (33,83.4521) (34,81.6263) (35,79.851) (36,83.219) (37,77.4574) (38,80.9946) (39,80.3158) (40,81.6239) (41,84.1033) (42,79.6423) (43,82.0302) (44,81.5597) (45,77.8667) (46,81.6593) (47,76.3478) (48,82.4496) (49,78.8139) (50,80.9075) (51,77.648) (52,79.9739) (53,78.9218) (54,79.044) (55,78.1741) (56,76.7406) (57,79.2917) (58,75.9345) (59,75.4598) (60,74.5887) (61,76.06) (62,73.824) (63,73.0688) (64,73.1153) };
    \addlegendentry{$n=100$ M, $1^+$-bit}
    \addplot+[colorPlotBlue] coordinates { (2,226.467) (3,243.5) (4,205.111) (5,217.9) (6,210.253) (7,188.756) (8,200.286) (9,200.858) (10,196.607) (11,197.46) (12,197.394) (13,200.167) (14,189.713) (15,192.643) (16,188.911) (17,186.2) (18,193.859) (19,194.733) (20,185.825) (21,193.213) (22,194.997) (23,192.939) (24,192.73) (25,191.636) (26,192.861) (27,185.777) (28,189.664) (29,191.433) (30,187.789) (31,188.518) (32,189.74) (33,191.148) (34,188.085) (35,189.841) (36,192.36) (37,189.863) (38,190.973) (39,192.018) (40,189.511) (41,190.323) (42,187.875) (43,189.514) (44,188.383) (45,187.309) (46,186.087) (47,183.065) (48,186.451) (49,182.118) (50,178.227) (51,181.332) (52,176.269) (53,175.508) (54,173.811) (55,173.999) (56,171.612) (57,168.24) (58,169.041) (59,165.609) (60,165.195) (61,165.36) (62,162.091) (63,161.614) (64,160.334) };
    \addlegendentry{$n=100$ M, 2-bit}
    \addplot+[colorPlotOrange] coordinates { (2,719.867) (3,158.267) (4,84.4889) (5,93.4) (6,73.9867) (7,157.189) (8,218.6) (9,155.317) (10,178.6) (11,168.26) (12,99.2545) (13,163.211) (14,151.246) (15,201.471) (16,175.387) (17,165.879) (18,152.753) (19,101.104) (20,119.34) (21,125.423) (22,136.365) (23,111.961) (24,140.017) (25,132.836) (26,112.843) (27,110.672) (28,111.909) (29,125.024) (30,132.0) (31,112.453) (32,124.178) (33,123.944) (34,112.206) (35,130.894) (36,128.16) (37,113.1) (38,118.184) (39,103.558) (40,123.441) (41,115.607) (42,126.985) (43,113.835) (44,117.705) (45,112.505) (46,131.298) (47,112.658) (48,114.638) (49,108.569) (50,98.5701) (51,114.839) (52,108.942) (53,105.336) (54,106.103) (55,100.228) (56,98.7842) (57,108.976) (58,111.924) (59,94.9713) (60,104.577) (61,104.742) (62,103.173) (63,109.817) (64,106.799) };
    \addlegendentry{$n=1000$ M, $1^+$-bit}
    \addplot+[colorPlotOrange] coordinates { (2,269.333) (3,267.133) (4,226.467) (5,290.633) (6,302.653) (7,297.122) (8,271.79) (9,286.958) (10,260.763) (11,272.713) (12,249.491) (13,271.478) (14,301.215) (15,271.443) (16,271.867) (17,293.642) (18,274.345) (19,283.57) (20,282.607) (21,267.617) (22,278.073) (23,298.676) (24,283.087) (25,280.378) (26,287.8) (27,280.513) (28,280.254) (29,284.605) (30,272.867) (31,272.318) (32,270.849) (33,267.596) (34,272.578) (35,261.28) (36,256.63) (37,257.809) (38,255.004) (39,253.461) (40,256.285) (41,249.238) (42,245.564) (43,251.522) (44,245.774) (45,245.721) (46,242.727) (47,240.796) (48,246.953) (49,239.486) (50,245.773) (51,238.289) (52,237.807) (53,234.964) (54,228.029) (55,235.127) (56,233.733) (57,233.146) (58,235.057) (59,228.495) (60,233.316) (61,233.701) (62,228.964) (63,225.408) (64,227.735) };
    \addlegendentry{$n=1000$ M, 2-bit}
    \end{axis}
\end{tikzpicture}

%% file: fig/combined1bit.tex
\begin{tikzpicture}
    \begin{axis}[
      xlabel={Bytes space overhead per additional thread},
      ylabel={Speedup},
      legend to name=legendCombined1,
      legend columns=2,
      xmin=0,
      ymin=0,
      width=5cm,
      height=3.5cm,
    ]
    \addplot+[colorPlotGreen] coordinates { (129.708,4.87921) };
    \addlegendentry{$n=10$ M, $1^+$-bit, minbpt=3}
    \addplot+[colorPlotGreen] coordinates { (43.6731,4.47334) };
    \addlegendentry{$n=10$ M, $1^+$-bit, minbpt=1000}
    \addplot+[colorPlotBlue] coordinates { (161.753,10.2028) };
    \addlegendentry{$n=100$ M, $1^+$-bit, minbpt=3}
    \addplot+[colorPlotBlue] coordinates { (85.3161,9.91666) };
    \addlegendentry{$n=100$ M, $1^+$-bit, minbpt=1000}
    \addplot+[colorPlotOrange] coordinates { (149.544,12.954) };
    \addlegendentry{$n=1000$ M, $1^+$-bit, minbpt=3}
    \addplot+[colorPlotOrange] coordinates { (124.178,12.848) };
    \addlegendentry{$n=1000$ M, $1^+$-bit, minbpt=1000}
    \end{axis}
\end{tikzpicture}

%% file: fig/combined2bit.tex
\begin{tikzpicture}
    \begin{axis}[
      xlabel={Bytes space overhead per additional thread},
      ylabel={Speedup},
      legend to name=legendCombined2,
      xmin=0,
      ymin=0,
      width=5cm,
      height=3.5cm,
    ]
    \addplot+[colorPlotGreen] coordinates { (325.6,4.66786) };
    \addlegendentry{$n=10$ M, 2-bit, minbpt=3}
    \addplot+[colorPlotGreen] coordinates { (103.701,4.46557) };
    \addlegendentry{$n=10$ M, 2-bit, minbpt=1000}
    \addplot+[colorPlotBlue] coordinates { (380.858,10.4134) };
    \addlegendentry{$n=100$ M, 2-bit, minbpt=3}
    \addplot+[colorPlotBlue] coordinates { (189.74,10.0774) };
    \addlegendentry{$n=100$ M, 2-bit, minbpt=1000}
    \addplot+[colorPlotOrange] coordinates { (382.271,12.8338) };
    \addlegendentry{$n=1000$ M, 2-bit, minbpt=3}
    \addplot+[colorPlotOrange] coordinates { (270.849,12.7931) };
    \addlegendentry{$n=1000$ M, 2-bit, minbpt=1000}
    \end{axis}
\end{tikzpicture}

%% file: fig/search.tex
\begin{tikzpicture}
    \begin{axis}[
      xlabel={Bytes space overhead per additional thread},
      ytick={1,2,3,4,5,6,7,8},
      yticklabels={{diff, $1^+$-bit},{maxprev, $1^+$-bit},{minbump, $1^+$-bit},{nosearch, $1^+$-bit},{diff, 2-bit},{maxprev, 2-bit},{minbump, 2-bit},{nosearch, 2-bit}},
      xbar,
      /pgf/bar width=8pt,
      nodes near coords,
      nodes near coords align={horizontal},
      enlarge x limits={value=0.2,upper},
      enlarge y limits=0.1,
      xmin=0,
      ymin=1,
      ymax=8,
      width=8cm,
      height=3.5cm,
    ]
    \addplot+[color=colorPlotBlue] coordinates { (106.596,1) (95.2839,2) (121.658,3) (128.228,4) (258.815,5) (258.544,6) (267.652,7) (273.996,8) };
    \end{axis}
\end{tikzpicture}

%% file: fig/speedup.tex
\begin{tikzpicture}
    \begin{axis}[
      xlabel={Threads},
      ylabel={Speedup},
      ylabel style={align=center},
      legend style={at={(1.1, 0.5)},anchor=west},
      width=8cm,
      height=3.5cm,
      mark repeat=4,
      xtick distance=8,
    ]
    \addplot+[colorPlotGreen] coordinates { (1,1.0) (2,1.60741) (3,2.10549) (4,2.5597) (5,3.07186) (6,3.28994) (7,3.48003) (8,3.66113) (9,3.89153) (10,3.99792) (11,4.10978) (12,4.2693) (13,4.33692) (14,4.38691) (15,4.5641) (16,4.62618) (17,4.6237) (18,4.73675) (19,4.72164) (20,4.74894) (21,4.84985) (22,4.71541) (23,4.48496) (24,4.67143) (25,4.68301) (26,4.59815) (27,4.59846) (28,4.69288) (29,4.28225) (30,4.64421) (31,4.15716) (32,4.47334) (33,4.1809) (34,4.06803) (35,4.03573) (36,4.13083) (37,3.83657) (38,3.93567) (39,3.84496) (40,3.84099) (41,3.95339) (42,3.76275) (43,3.85688) (44,3.60644) (45,3.68683) (46,3.54412) (47,3.70485) (48,3.61033) (49,3.39444) (50,3.40115) (51,3.54824) (52,3.52635) (53,3.38755) (54,3.44817) (55,3.84391) (56,3.72792) (57,3.76872) (58,3.58946) (59,3.60481) (60,3.65783) (61,3.65758) (62,3.62136) (63,3.5003) (64,3.35159) };
    \addlegendentry{$n=10$ M, $1^+$-bit}
    \addplot+[colorPlotGreen] coordinates { (1,1.0) (2,1.65109) (3,2.15979) (4,2.61089) (5,3.07345) (6,3.27609) (7,3.47534) (8,3.6821) (9,3.86946) (10,4.00593) (11,4.09772) (12,4.29562) (13,4.29511) (14,4.44641) (15,4.56572) (16,4.63075) (17,4.61866) (18,4.7496) (19,4.78857) (20,4.85541) (21,4.74964) (22,4.77426) (23,4.83044) (24,4.55366) (25,4.61849) (26,4.76389) (27,4.68593) (28,4.70286) (29,4.58003) (30,4.38034) (31,4.51601) (32,4.46557) (33,4.09734) (34,4.36705) (35,3.9612) (36,4.00238) (37,4.08908) (38,4.12091) (39,4.51012) (40,4.40486) (41,4.27219) (42,4.24279) (43,4.27573) (44,4.03474) (45,4.10472) (46,4.23547) (47,4.0904) (48,3.78447) (49,4.05742) (50,4.00815) (51,4.05243) (52,3.97848) (53,3.72815) (54,3.6396) (55,3.64508) (56,3.70519) (57,3.57409) (58,3.62064) (59,3.52535) (60,3.47584) (61,3.52974) (62,3.57641) (63,3.42501) (64,3.54608) };
    \addlegendentry{$n=10$ M, 2-bit}
    \addplot+[colorPlotBlue] coordinates { (1,1.0) (2,1.86485) (3,2.6515) (4,3.30955) (5,3.9862) (6,4.61131) (7,5.13957) (8,5.64882) (9,6.07676) (10,6.40519) (11,6.71739) (12,6.99144) (13,6.96367) (14,7.20399) (15,7.41524) (16,7.78395) (17,7.91697) (18,8.0948) (19,8.3082) (20,8.50462) (21,8.67033) (22,8.79753) (23,9.02802) (24,9.05798) (25,9.29767) (26,9.36393) (27,9.45688) (28,9.53716) (29,9.66374) (30,9.66817) (31,9.69717) (32,9.91666) (33,9.47281) (34,9.62221) (35,9.5704) (36,9.64897) (37,9.59418) (38,9.70003) (39,9.58883) (40,9.76955) (41,10.0334) (42,10.1779) (43,10.18) (44,10.2802) (45,10.2379) (46,10.3087) (47,10.2026) (48,10.4279) (49,10.2774) (50,10.5714) (51,10.4531) (52,10.5243) (53,10.3983) (54,10.5944) (55,10.5778) (56,10.5085) (57,10.4946) (58,10.7063) (59,10.5679) (60,10.5639) (61,10.446) (62,10.4827) (63,10.3888) (64,10.3784) };
    \addlegendentry{$n=100$ M, $1^+$-bit}
    \addplot+[colorPlotBlue] coordinates { (1,1.0) (2,1.87533) (3,2.66754) (4,3.35314) (5,4.0013) (6,4.61874) (7,5.18484) (8,5.67606) (9,6.03667) (10,6.44928) (11,6.78159) (12,7.07098) (13,6.98921) (14,7.22576) (15,7.46592) (16,7.74322) (17,7.88065) (18,8.11519) (19,8.25906) (20,8.46277) (21,8.64695) (22,8.78455) (23,9.02492) (24,9.05892) (25,9.39823) (26,9.403) (27,9.53632) (28,9.60254) (29,9.76213) (30,9.83536) (31,10.0061) (32,10.0774) (33,9.66422) (34,9.87405) (35,9.65543) (36,9.7806) (37,9.83749) (38,9.91497) (39,9.91015) (40,10.0627) (41,10.1498) (42,10.1618) (43,10.0842) (44,10.0556) (45,10.35) (46,10.1872) (47,10.3059) (48,10.3909) (49,10.3884) (50,10.3216) (51,10.3324) (52,10.3742) (53,10.3868) (54,10.4695) (55,10.3914) (56,10.4268) (57,10.3511) (58,10.3194) (59,10.4287) (60,10.3847) (61,10.355) (62,10.3854) (63,10.4445) (64,10.4632) };
    \addlegendentry{$n=100$ M, 2-bit}
    \addplot+[colorPlotOrange] coordinates { (1,1.0) (2,1.86058) (3,2.67852) (4,3.30799) (5,4.05017) (6,4.76036) (7,5.36868) (8,5.90947) (9,6.4586) (10,7.02) (11,7.54723) (12,7.74529) (13,7.67306) (14,8.12474) (15,8.50699) (16,8.92433) (17,9.32092) (18,9.62898) (19,9.93705) (20,10.2363) (21,10.5143) (22,10.7567) (23,10.9832) (24,11.2648) (25,11.5011) (26,11.7106) (27,11.914) (28,12.0789) (29,12.2527) (30,12.4934) (31,12.6395) (32,12.848) (33,12.4598) (34,12.5057) (35,12.7118) (36,12.8425) (37,12.8093) (38,12.9281) (39,13.0693) (40,13.151) (41,13.2659) (42,13.327) (43,13.4593) (44,13.3962) (45,13.5808) (46,13.6482) (47,13.6349) (48,13.7479) (49,13.76) (50,13.8407) (51,13.8505) (52,13.8643) (53,13.9588) (54,13.9735) (55,14.0688) (56,14.0522) (57,14.1137) (58,14.1812) (59,14.2415) (60,14.2295) (61,14.318) (62,14.3745) (63,14.3598) (64,14.2986) };
    \addlegendentry{$n=1000$ M, $1^+$-bit}
    \addplot+[colorPlotOrange] coordinates { (1,1.0) (2,1.86472) (3,2.67538) (4,3.30692) (5,4.02791) (6,4.77516) (7,5.41739) (8,5.99484) (9,6.50737) (10,7.017) (11,7.51631) (12,7.87639) (13,7.77463) (14,8.11247) (15,8.53301) (16,8.98812) (17,9.27434) (18,9.67112) (19,9.9934) (20,10.2048) (21,10.5367) (22,10.7588) (23,11.07) (24,11.2876) (25,11.5468) (26,11.7145) (27,11.9267) (28,12.1014) (29,12.2816) (30,12.4449) (31,12.6072) (32,12.7931) (33,12.4045) (34,12.5333) (35,12.6496) (36,12.8505) (37,12.8139) (38,12.905) (39,13.0466) (40,13.082) (41,13.1819) (42,13.2853) (43,13.2948) (44,13.4634) (45,13.5411) (46,13.5516) (47,13.618) (48,13.8049) (49,13.674) (50,13.8579) (51,13.8253) (52,13.8584) (53,13.8529) (54,13.9473) (55,13.9186) (56,13.9893) (57,14.0092) (58,14.1627) (59,14.2095) (60,14.2578) (61,14.1461) (62,14.3664) (63,14.3139) (64,14.2565) };
    \addlegendentry{$n=1000$ M, 2-bit}

    \addplot[color=gray,dashed] coordinates { (32,1) (32,15) };
    \node[color=gray] at (axis cs: 34,14.5) {\tiny HT};
    \end{axis}
\end{tikzpicture}

%% file: fig/breakdown.tex
\begin{tikzpicture}
    \begin{axis}[
      title style={align=center},
      xlabel={Threads},
      ylabel={Time (s)},
      xtick=data,
      xticklabels={1,2,4,8,16,32,64},
      ybar,
      enlarge x limits={value=0.2,upper},
      xmin=0,
      scaled y ticks=false,
      x=1.5cm,
      width=8cm,
      height=3.5cm,
      no marks,
    ]
    \addplot[fill,colorPlotGreen] coordinates { (1,10.5172) (2,5.55484) (3,3.02119) (4,1.75678) (5,1.24197) (6,0.862146) (7,0.694406) };
    \addlegendentry{Insertion};
    \addplot[fill,colorPlotBlue] coordinates { (1,3.41245) (2,1.76892) (3,1.09628) (4,0.643914) (5,0.49879) (6,0.484368) (7,0.608552) };
    \addlegendentry{Sorting};
    \addplot[fill,colorPlotOrange] coordinates { (1,1.49996) (2,0.903896) (3,0.484065) (4,0.317671) (5,0.251901) (6,0.184591) (7,0.171693) };
    \addlegendentry{Back Substitution};
    \end{axis}
\end{tikzpicture}